# Growth mechanism of polymer membranes obtained by H-bonding across immiscible liquid interfaces

J. Dupré de Baubigny[1], P. Perrin[1], N. Pantoustier[1], T. Salez*[2,3], M. Reyssat[4] and C. Monteux*[1,3]

1. Sciences et Ingénierie de La Matière Molle, ESPCI Paris, PSL Research University, CNRS, Sorbonne Universités, UPMC Univ Paris 06, 75005 Paris, France
2. Univ. Bordeaux, CNRS, LOMA, UMR 5798, 33405 Talence, France
3. Global Station for Soft Matter, Global Institution for Collaborative Research and Education, Hokkaido University, Sapporo, Japan
4. UMR CNRS 7083 Gulliver, ESPCI Paris, PSL Research university, 75005 Paris, France

* corresponding authors: cecile.monteux@espci.fr and thomas.salez@u-bordeaux.fr

**ABSTRACT**

Complexation of polymers at liquid interfaces is an emerging technique to produce all-liquid printable and self-healing devices and membranes. It is crucial to control the assembly process but the mechanisms at play remain unclear. Using two different reflectometric methods, we investigate the spontaneous growth of H-bonded PPO-PMAA (Poly Propylene oxide- Poly metacrylic acid) membranes at a flat liquid-liquid interface. We find that the membrane thickness h grows with time t as $h \sim t^{1/2}$, which is reminiscent of a diffusion-limited process. However, counter-intuitively, we observe that this process is faster as the PPO molar mass increases. We are able to rationalize these results with a model which considers the diffusion of the PPO chains within the growing membrane. The architecture of the latter is described as a gel-like porous network, with a pore size much smaller than the radius of the diffusing PPO chains, thus inducing entropic barriers that hinder the diffusion process. From the comparison between the experimental data and the result of the model, we extract some key piece of information about the microscopic structure of the membrane. This study opens the route toward the rational design of self-assembled membranes and capsules with optimal properties.

Complexation of polymers, surfactants or particles at immiscible liquid interfaces is an increasingly popular technique to produce all-liquid printable, reconfigurable and self-healing membranes, devices and capsules[1–12]. While layer-by-layer assembly of components at liquid interfaces enables to obtain good control over membrane thickness and composition[13–15], an easier way to promote interfacial complexation is to dissolve the interacting species within two separate liquid phases, such as oil and water[6,9–11,16–21] or two partially-miscible phases [2,5,22–24]. As the components spontaneously diffuse towards the common interface, they self-assemble through non-covalent interactions such as electrostatic ones and form a membrane which grows over time up to micrometric thicknesses[10,11]. However, the current lack of understanding of the mechanisms at play during the assembly process hinders the development of membranes with controlled structure and properties. Indeed the very few experimental results available concerning the kinetics of growth of self-assembled membranes[10,25] do not provide a clear microscopic picture of the assembly process. Capito *et al.*[2] who was the first to assemble membranes using peptides and polysaccharides of opposite charge, assumed that the membrane growth was controlled by the diffusion of small peptides through the growing peptide-polysaccharide membrane. However the results obtained later by Mendoza-Meinhardt *et al.*[25] for a similar system[5] where inconsistent with a diffusion-limited process. The diffusion of molecules in polymer networks has been the object of a large amount of theoretical and experimental studies, but is still an unsolved question[26–33]. In this Letter, we use interferometric *in-situ* and *ex-situ* measurements to follow the thickness evolution of a model self-assembled interfacial membrane obtained from the H-bond complexation of poly(methacrylic acid) (PMAA) and poly(propylene oxide) (PPO) at a flat isopropylmyristate (IPM)-water interface. We showed recently that this system enables to obtain highly-stable oil-water emulsions using a simple rotor-stator emulsification technique[19]. We measure the membrane thickness as a function of time during its spontaneous complexation, for various polymer concentrations and molar masses. We find that the assembly process is diffusion-limited, controlled by both the PPO concentration and molar mass. Our measurements show that the diffusion coefficients of the PPO chains in the growing membrane are extremely slow. To account for these results, we suggest that the diffusion of free PPO chains within the growing membrane is hindered by entropic barriers due to the low mesh size of the polymer network. A minimal model including this assumption enables us to rationalize all the macroscopic data and extract some key information about the microscopic structure of the PPO-PMAA membrane thus opening the way towards its optimal design.

To obtain insight into the mechanisms at play during the growth of polymer membranes at immiscible liquid interfaces we choose to work with two polymers that interact through hydrogen bonds, poly(methacrylic acid) (PMAA) as a H-bond donor and poly(propylene oxide) (PPO) as a H-bond acceptor (Figures 1a and 1b). We dissolve both polymers in two immiscible phases, the PPO in isopropylmyristate (IPM) and the PMAA in water. When the two polymer phases are put into contact in a container a membrane instantaneously forms at the flat IPM/water interface (Figure 1c), which thickness time evolution is measured using two different interferometric methods. Briefly we perform an *in situ* measurement at a flat IPM/water interface using an optical spectrometer (Specim V8E) to measure the wavelength dependence of the light intensity reflected by the thin interfacial membrane and an *ex-situ* method consisting in removing the membrane from the liquid and leaving it on a glass slide and measuring its thickness with an optical interferometric profilometer (Microsurf 3D Fogale Nanotech).

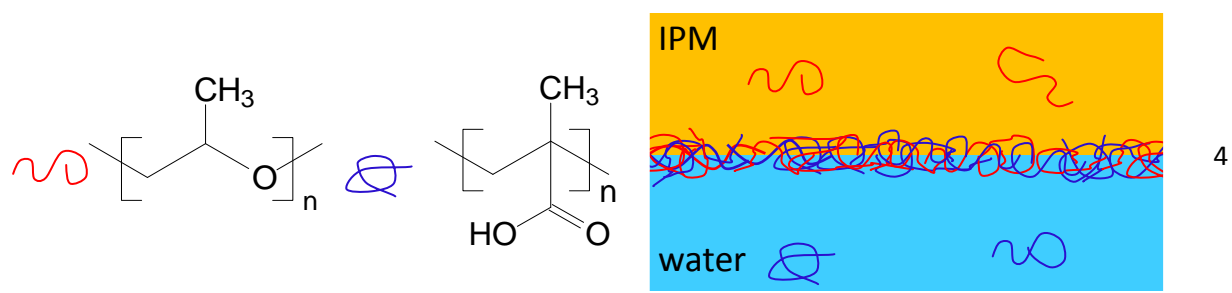

1a  1b  1c

*Figure 1. Chemical formulas of a. Poly(Propylene Oxide) (PPO) and b. Poly(Methacrylic Acid) (PMAA) c. Schematic diagram of the interfacial complexation between PMAA (dark blue) and PPO (red) at the water (blue)-IPM (orange) interface, leading to the membrane formation.*

In Figure 2a, we quantitatively investigate the growth of the PPO-PMAA membrane at the flat IPM-water interface, for $M_{n,PPO}$ = 3000 g/mol and $M_{n,PMAA}$ = 100 000 g/mol (Figure 2a) at varying weight fractions. In Figure 2b, we report on the membrane growth for several PPO and PMAA molar masses with a weight percentage fixed to 1 wt% for both polymers. For all the experimental conditions, the membrane thickness $h(t)$ is found to scale with time $t$ as $h \sim t^{1/2}$, over up to four decades in time, which suggests the existence of an underlying diffusive process characterized by a diffusion coefficient that remains constant over the course of each experiment. These results differ from those presented in the works of Gazizov[23] and Mendoza-Meinhardt[25], where the thickness growth was observed to reach a plateau after a few minutes or hours respectively. The effective diffusion coefficients $D_{eff}$ obtained from the experimental curves using a fit of the form $h(t) = (2 D_{eff} t)^{1/2}$, range between $10^{-19}$ m$^2$/s and $10^{-17}$ m$^2$/s depending on the experimental conditions (Table 1, third column). These values are six to eight orders of magnitude lower than the bulk diffusion coefficients expected for PPO in IPM or PMAA in water. Extremely low effective diffusion coefficients were also measured by Gunes *et al.* for a chitosan-phospholipid membrane growing at an oil-water interface[10]. These low values of $D_{eff}$ are inconsistent with the free diffusion of the polymers in the bulk phases and we therefore suggest that the diffusion of the polymers through the growing membrane controls the assembly process. Indeed, the H-bond complexation between PMAA and PPO at the water-IPM interface is fast, and a thin membrane is almost instantaneously formed as soon as the two phases are put in contact. To induce further membrane growth, it is then reasonable to expect that the PPO chains (respectively the PMAA chains) in the organic IPM phase (resp. the aqueous phase) have to diffuse through the membrane to reach the aqueous phase (resp. the organic phase) where they can complex with PMAA (resp. PPO).

2a

2b

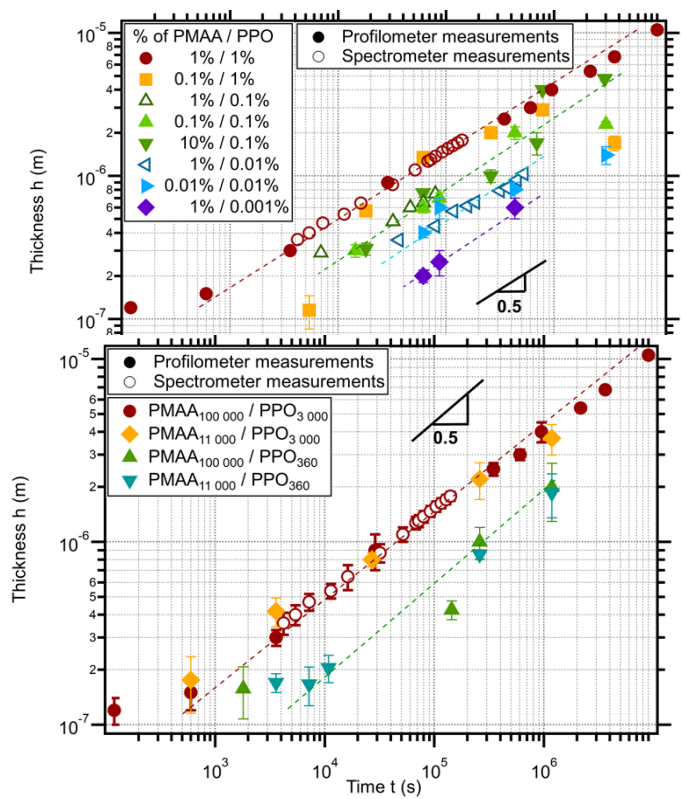

*Figure 2. a. Thickness of the PPO-PMAA membrane at the IPM-water interface as a function of time, for varying PMAA and PPO weight percentages between 0.001 wt% and 1 wt%, as measured either with a profilometer (filled symbols) or with a spectrometer (open symbols). The molar masses are: $M_{n,PPO} = 3000$ g/mol and $M_{n,PMAA} = 100\,000$ g/mol. b. Thickness of the PPO-PMAA membrane at the IPM-water interface for 1 wt% of PMAA and 1wt% of PPO, and varying number-averaged molar masses: $M_{n,PPO} = 3000$ g/mol or 360 g/mol, and $M_{n,PMAA} = 11000$ g/mol or 100000 g/mol.*

We further see from Figure 2a that increasing the bulk PMAA concentration in the aqueous phase does not influence the membrane growth, while increasing the PPO concentration in the organic phase leads to thicker membranes. Consistently, the values of the effective diffusion coefficient, $D_{\text{eff}}$, obtained from Figure 2a and reported in Table 1 (fourth column) increase with the bulk PPO concentration. This suggests that only the PPO chains diffuse through the PPO-PMAA membrane before reaching the aqueous phase. Consistently we find that the PMAA molar mass does not influence the growth of the membrane (Figure 2b). However increasing the PPO molar mass leads to thicker membranes and to a higher value of $D_{\text{eff}}$, which may seem counter-intuitive at first sight as diffusion coefficients of macromolecules are expected to decrease with molar mass.

|  | wt % | $C_{\text{bulk}}^{\text{PPO}}$ x10$^{24}$ molecules/m$^3$ | $D_{\text{eff}}$ x10$^{-17}$ m$^2$/s | $D_{\text{m}}^{\text{PPO}}$ x10$^{-15}$ m$^2$/s | $\xi$ nm |
|---|---|---|---|---|---|
| $M_{\text{n,PPO}} = 3000$ g/mol $N_{\text{PPO}} = 52$ | 0.001 | 0.002 | 0.04 | 4.55 | 0.59 |
|  | 0.01 | 0.02 | 0.14 | 1.70 | 0.55 |
|  | 0.1 | 0.2 | 0.33 | 0.39 | 0.51 |
|  | 1 | 2.0 | 1.4 | 0.17 | 0.48 |
| $M_{\text{n,PPO}} = 360$ g/mol $N_{\text{PPO}} = 6$ | 1 | 17 | 0.15 | 0.11 | 0.12 |

*Table 1. Effective and microscopic diffusion coefficients of PPO chains inside the membrane, $D_{\text{eff}}$ (Equation 2) and $D_{\text{m}}^{\text{PPO}}$(Equation 3) respectively, as obtained from Figure 2. The last*

*column represents the pore size ξ(nm) of the PPO-PMAA membrane estimated from Equation 5, using the best-fit parameters from the comparison between all the data and Equation 6.*

To summarize, our experimental results show that the growth kinetics of the membrane is controlled by both the PPO molecular mass and concentration, and that the PMAA molar mass and concentration do not influence the process. These experimental results suggest that the membrane-assembly process is limited by the diffusion of the PPO chains through the growing membrane. In the following, we will therefore consider the diffusive transport of PPO chains through a membrane of thickness $h(t)$ along the $z$ direction, and composed of complexed PMAA and PPO chains (Figure 3a). At the membrane-water interface ($z = h(t)$), the free PPO chains interact and complex with the free PMAA chains, leading to membrane growth. This implies a smaller free PPO concentration $C_{z=h(t)}^{\text{PPO}}$ at the membrane-water interface with respect to the bulk value $C_{\text{bulk}}^{\text{PPO}}$, and thus the existence of a gradient of PPO molecular concentration $C^{\text{PPO}}$ across the membrane. The molecular flux along $z$ of PPO chains diffusing through the membrane from the IPM phase to the water phase thus reads $J = -D_{\text{m}}^{\text{PPO}} \frac{\partial C^{\text{PPO}}}{\partial z}$, with $D_m^{PPO}$ the diffusion coefficient of the free PPO chains in the membrane. Assuming an almost instantaneous complexation at the membrane-water interface, we can further write $C_{z=h(t)}^{\text{PPO}} \ll C_{\text{bulk}}^{\text{PPO}}$, and consider the membrane growth as a relatively slow, diffusion-limited process with a quasi-steady concentration profile $C^{\text{PPO}}(z)$. The diffusive flux across the membrane therefore becomes $J \approx D_{\text{m}}^{\text{PPO}} \frac{C_{\text{bulk}}^{\text{PPO}}}{h(t)}$, and the governing equation for the evolution of the membrane thickness is:

$$\frac{dh}{dt} \approx vJ \approx v D_{\text{m}}^{\text{PPO}} \frac{C_{\text{bulk}}^{\text{PPO}}}{h(t)} , \qquad (1)$$

with $v$ the volume of a PPO molecule.

Equation 1 can be integrated, under the $h(0) = 0$ initial condition, and the solution reads:

$$h(t) \approx (2D_{\text{eff}}^{\text{PPO}} t)^{1/2} , \qquad (2)$$

with $D_{\text{eff}}^{\text{PPO}} = v C_{\text{bulk}}^{\text{PPO}} D_{\text{m}}^{\text{PPO}}$ . $\qquad (3)$

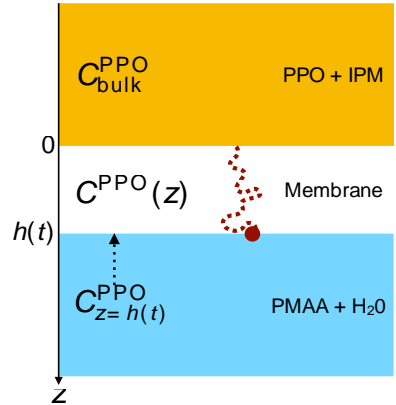

3a

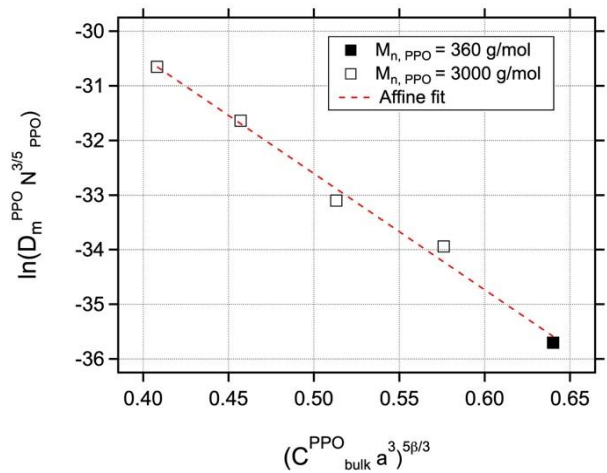

3b

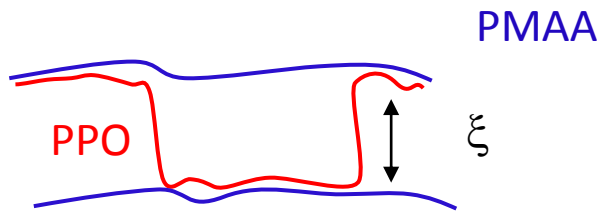

3c

*Figure 3. a. Schematic showing the diffusion (red dotted trajectory) of a PPO molecule (red disk) inside a PPO-PMAA membrane (white), from a bulk PPO solution in IPM (z = 0, orange) to the interface with a bulk PMMA solution in water (z = h(t), blue) where PPO-PMAA complexation occurs. The respective PPO concentrations are indicated. b. Logarithm*

*of the rescaled diffusion coefficient of PPO molecules in the membrane as a function of the rescaled bulk PPO concentration to the power 5β/3, with β=0.03. The equation of the affine fit is: -21.3 ($C^{PPO}a^3$)$^{0.05}$-22. 3c. Schematic showing the possible structure of the PPO/PMAA chains in the interfacial membrane*

Using Equation 3 and the values of $D_{eff}$ reported in Table 1, we can deduce the corresponding microscopic diffusion coefficients $D_m^{PPO}$ of a PPO chain inside the membrane for the various experimental conditions (Table 1 fifth column). To estimate the volume $v$ of a diffusing PPO molecule, we assume that it is in dilute and athermal-solvent conditions inside the membrane, which is a reasonable assumption considering that traces of IPM are likely to be present in the membrane. As such, their Flory radius is given by $R_F \approx a N_{PPO}^{3/5}$, with $a \approx 0,2$ nm the monomeric size, $N_{PPO}$ the number of monomers per PPO chain, and thus $v \approx \frac{4}{3}\pi R_F^3$. The obtained values for $D_m^{PPO}$ shown in Table 1 range between $10^{-16}$ and $4.5 \cdot 10^{-15}$ m²/s, which are five to four orders of magnitude lower than the free diffusion coefficient of diluted PPO chains in a pure solvent. Moreover, Table 1 shows that the diffusion coefficient $D_m^{PPO}$ of PPO decreases as the bulk PPO concentration increases.

To understand further the values and dependencies of the microscopic diffusion coefficient $D_m^{PPO}$, we have to take into account the fact that the free PPO chains diffuse in a complex polymer network constituting the membrane. Moreover, another difficulty arises from the fact that the microscopic structure of the latter is unknown and remains an open question in the literature. Nevertheless, one can invoke a few minimal intuitive assumptions. First of all, the solid PPO-PMAA matrix of the membrane is likely to be a porous network, with a typical pore size $\xi$ that can *a priori* depend on the PPO concentration and molar mass. Secondly, we suggest that the polymer matrix is dense, *i.e.* $\xi$ is small compared to the Flory radius $R_F$ of the free PPO molecules. This implies the existence of entropic barriers hindering the diffusion of the PPO molecules within the membrane[32]. Thirdly, multiplying the Zimm bulk diffusion coefficient for real chains diluted in a good solvent, to an Arrhenius factor involving an elastic barrier of entropic origin, one arrives at the following expression:

$$D_{\mathrm{m}}^{\mathrm{PPO}} \approx \frac{D_0}{N_{\mathrm{PPO}}^{3/5}} \exp\left[-\left(\frac{R_F}{\xi}\right)^{5/3}\right] \quad , \qquad (4)$$

with $D_0 = \frac{kT}{6\pi\eta a}$ the monomeric Stokes-Einstein diffusion coefficient in a medium of viscosity $\eta$, at temperature $T = 293$ K, and $k$ the Boltzmann constant. Finally, since the membrane is composed of short PPO chains bridging long PMAA chains together, we might expect that the typical pore size $\xi$ scales like the PPO size, $R_F \approx aN_{\mathrm{PPO}}^{3/5}$, with a concentration-dependent correction prefactor accounting for the actual fraction of PPO crosslinkers. Assuming the corrective prefactor to be a power law with an unknown exponent $-\beta$, one gets the Ansatz:

$$\xi \approx a\lambda N_{PPO}^{3/5}\left(C_{bulk}^{PPO}a^3\right)^{-\beta}, \quad (5)$$

where $\lambda$ is a dimensionless numerical constant. Combining Equations 4 and 5, we finally obtain the following prediction:

$$D_{\mathrm{m}}^{\mathrm{PPO}} \approx \frac{D_0}{N_{\mathrm{PPO}}^{3/5}} \exp\left[-\frac{(C_{\mathrm{bulk}}^{PPO}a^3)^{\frac{5\beta}{3}})}{\lambda^{\frac{5}{3}}}\right] . \quad (6)$$

By fitting the data of Table 1 to Equation 6, one gets a good agreement as shown in Figure 3b, obtained for $\beta = 0.03$. From the affine fit, we deduce the values: $\lambda = 0.16$ and $D_0 = 2.8\ 10^{-10}$ m$^2$/s. The value of $\xi$ calculated from Equation 5 (Table 1, last column), using those best-fit parameters, ranges between $5\ 10^{-10}$ and $6\ 10^{-10}$ m for the large polymer chains ($R_F = 25\ 10^{-10}$ m), and is around $10^{-10}$ m for the small polymer chains ($R_F = 6.4\ 10^{-10}$ m), which are all self-consistently smaller than the PPO molecular size. We note that the values of $\xi$ are on the same order of magnitude as the persistence length expected for both polymers[34–37]. We also stress that the value of $D_0$ is comparable to the free diffusion coefficient of a monomer in a pure solvent, on the order of $10^{-10}$ m$^2$/s.

Increasing the bulk PPO concentration, and thus the crosslinking fraction in the membrane, leads to a decrease of the average pore size, as expected. Indeed, a NRM measurement (not shown) of the membrane composition showed that the polymer stoichiometry of the membrane is identical to the bulk stoichiometry. As such, increasing the PPO concentration in the IPM phase leads to an increased concentration of PPO in the membrane, and thus a decrease of the pore size. Furthermore, the above values of $\xi$ are consistent with a microscopic membrane architecture where the PPO chains act as macromolecular cross-linkers with PPO monomeric units sticking to the PMAA chains and bridging them together in a zip-like fashion. In this picture, the unbounded PPO units constitute the porous network of the membrane (see Figure 3c). To confirm independently these low values of $\xi$, we invoke interfacial-rheology measurements[19] previously reported in our group for the shear elastic modulus of these membranes, on the order of 13 MPa. The shear elastic modulus, which scales as $G \approx kT/\xi^3$, enables us to estimate that $\xi$ is on the order of $5 \cdot 10^{-10}$ m, consistently with the values obtained above. Altogether, from this minimal model, one is able to discuss the microscopic structure of the membrane based on simple macroscopic measurements only.

## CONCLUSION

Using spectrometry and optical profilometry, we measure the growth kinetics of a PPO-PMAA membrane, resulting from H-bonding complexation at the water-Isopropyl Myristate interface. We find that the thickness of the membrane scales as the square root of time, consistently with a diffusive process. Systematic measurements for varying PPO and PMAA molar masses and concentrations lead us to the conclusion that the complexation process is limited by the diffusion of the PPO chains through the growing membrane. From the macroscopic growth curves we obtain the diffusion coefficient of the PPO chains in the membrane and find that it decreases with the PPO concentration and increases with the PPO molar mass. To rationalize these counter-intuitive observations, we model the process by considering that the growing membrane is a gel-like porous network, with a pore size smaller than the radius of the diffusing PPO chains, thus inducing entropic barriers which hinder the macromolecular diffusion. Moreover, we consider that the pore size of the membrane depends on the PPO concentration and molar mass. By fitting the experimental diffusion coefficients with the model prediction, we are able to deduce that the pore size of the growing membrane decreases with the PPO concentration and increases with the PPO molar mass, consistently with a membrane structure where the PMAA chains are bridged in a zip-like fashion by PPO molecules acting as macromolecular cross-linkers. Our model therefore enables us to discuss the microscopic structure of the PPO-PMAA membrane from macroscopic measurements only. This study opens the route toward a rational design of self-assembled membranes and capsules with optimal properties.

**EXPERIMENTAL SECTION**

The aqueous PMAA solutions are prepared by dissolving 0.001 to 1 wt% of PMAA in purified water from a milli-Q apparatus (Millipore) and the $pH$ is adjusted to $pH = 3$ by adding drops of a HCl (Sigma-Aldrich) solution concentrated at 1 mol/L, and measured with a $pH$-meter ($pH$ M 250 ion analyser, Meterlab, Radiometer Copenhagen). We used two PMAA samples, sold as PMAA 100 K from Polysciences and PMAA 10K from Sigma Aldrich. Molar masses and polydispersity indexes (PDI) were measured on a Size Exclusion

Chromatography (SEC, Viscotek GPC max VE 2001; TDA 302 triple detector array, system with triple detection using 0.2 M NaNO3 aqueous solution or THF containing 2% trimethylamine for PMAA and PPO samples respectively). Using SEC in aqueous solution, we measure for PMAA 100K a number averaged molar mass $M_n$ = 100000 g/mol and *PDI* around 2. PMAA 10K is purchased under the form of a sodium salt solution at a concentration of 30 wt%, which is then dialyzed against water at pH=3 for four weeks to obtain it under a PMAAH form and is then lyophilized. Using SEC in aqueous solution we measure $M_n$ = 11000 g/mol and *PDI* = 1.82. The oil-based solution is prepared by dissolving 0.001 to 1 wt% of PPO in Isopropyl Myristate, IPM (Sigma-Aldrich). We use two PPO from Sigma-Aldrich with product names PPO 400 and PPO 4000, which are used without any further purification. Using SEC in THF we find that PPO 4000 has a number averaged molar mass $M_n$=3000 and *PDI* = 1.97. We note that PPO with higher molar masses are not available with common chemical suppliers. As SEC is not suitable for short chains such as PPO 400, we analyzed it using GC-MS measurements. The measurements lead a molar mass of 360 g/mol. Despite the fact that this technique does not allow for the determination of a molecular weight distribution it can however can be reasonably assumed that $M_w = M_n$. We therefore assume that $M_w = M_n$ = 360 g/mol.

The interfacial complexation and the membrane are obtained by putting the two polymer phases in contact (Figure 1c). The membrane thickness is measured using two different methods. The first method consists in an *in situ* measurement of the thickness of the membrane assembled at a flat IPM/water interface. Briefly, we use a reflected light microscope mounted with an optical spectrometer (specim V8E) connected to a camera. We focus white light on the oil-water interface where the membrane grows and the spectrometer provides the wavelength dependent intensity reflected by the membrane from which we deduce the membrane thickness. We note that this technique is not suitable for thicknesses below 300 nm, for which it is difficult to detect the interference fringes. The second method consists in removing the membrane from the liquid and leaving it on a glass slide and measuring its thickness *ex situ*, with an optical interferometric profilometer (Microsurf 3D Fogale Nanotech).

**ACKNOWLEDGEMENTS**

The authors acknowledge the ANR JCJC INTERPOL grant number ANR-12-JS08-0007 for funding. They also acknowledge Alesya Mikhailovskaya and Corentin Trégouët for fruitful discussions, Julie Brun and Charlotte Demonsant for their help with the thickness measurements, as well as Mohamed Hanafi for performing the SEC measurements of the polymers.